\documentclass[amssymb,aps,twocolumn,showpacs,nofootinbib]{revtex4}

\usepackage[]{graphicx}
\usepackage{hyperref}

\def\ket#1{| #1 \rangle}

\def\bra#1{\langle #1 |}

\begin{document}

\title{Robust Quantum Communication Using A Polarization-Entangled Photon Pair}

\author{J.-C. Boileau, R. Laflamme,  M. Laforest, C. R. Myers}

\affiliation{Institute for Quantum Computing, University of Waterloo,
  Waterloo, ON, N2L 3G1,
  Canada.}

\date{\today}

\begin{abstract}
  Noise and imperfection of realistic devices are major obstacles for
  implementing quantum cryptography.  In particular, birefringence in
  optical fibres leads to decoherence of qubits encoded
  in photon polarization. We show how to overcome this problem by doing
  single qubit quantum communication without a shared spatial
  reference frame and precise timing. Quantum information will be
  encoded in pairs of photons using ``tag'' operations which
  corresponds to the time delay of one of the polarization modes.
  This method is robust against the phase instability of the
  interferometers despite the use of time-bins.  Moreover synchronized
  clocks are not required in the ideal no photon loss case
  as they are only necessary to label the different encoded qubits.
\end{abstract}


\pacs{03.67.Pp, 03.67.Hk,
  03.67.Dd}

\maketitle

Quantum mechanics allows the distribution of cryptographic keys whose
security is based on the laws of physics instead of the difficulty of
solving mathematical problems \cite{qc1983, BB84}.  Turning this idea
into practical technologies brings exciting challenges.  The first
prototype for quantum cryptography was built more than ten years ago
over a distance of 30 cm in free space \cite{BBBSJ1992} and used the photons' polarization as qubits of information.  Since, many quantum key distribution (QKD) experiments
have been realized through air and optic fibres \cite{GRTZ2002}. One
of the obstacles to improve the fibre based prototypes is the
birefringence effects due to geometric asymmetries and tension
fluctuations which are a major impediment for polarization based-coding
experiments \cite{GBGGHRTZ2000}.  When the coherence time of the photon
is large compared to the delay caused by polarization mode dispersion,
the birefringence can be represented by a time dependent unitary
transformation $U(t)$ that acts on the polarization space. The time
dependance comes from the mechanical variations in the fibre over time
and its rate varies with the environmental conditions.

A possible solution to this problem is the application of active
feedback \cite{FJ1995}. Tomography on some predetermined polarization
states could be used to approximate $U$ for a certain time interval \cite{J1991, RG2003}. By applying his approximation of $U^\dagger$ before his measurements, Bob (the receiver) could recover the states sent by Alice (the sender). However, this technique is practical only if the rate of change of $U$ is relatively low. For this reason, the most successful QKD experiments were not based on polarization coding, such as the phase based experiment proposed by Bennett \textit{et al.} using unbalanced interferometer \cite{Bennett1992, TRT1993, HMP2000}. However, a good control of the polarization modes is necessary to obtain a better visibility since some components like phase modulators are polarization dependent and the temperature of the interferometers must be stabilized since very small fluctuations between the two arms cause phase shifts that corrupt the quantum states. 

Another very important example of a successful QKD protocol is the
plug-and-play set-up \cite{MHHTZG1997, SGGRZ2002}.
Using a Faraday mirror \cite{Martinelli1989}, the photons sent by Bob are reflected back in the fibre by Alice, who in turn encodes information in their phase. By
travelling back in the fibre, the birefringence is reversed and, as it can be shown, the polarization state received by Bob are orthogonal to the original one. Since Bob controls the polarization state of the photon, he can make use of a polarized beamsplitter which increases the interference visibility. 
Although the plug-and-play set-up has very interesting
characteristics, it is not compatible with a non-Poissonian source
which could get rid of the multi-photons per pulse problem. Another
disadvantage is that the use of two-way quantum cryptography
is more vulnerable to a certain kind of eavesdropping strategy: the Trojan attack. An eavesdropper (i.e. Eve) could send photons in Alice's lab, catch them after they were reflected by the Faraday mirror and get some
information about Alice's set-up without being detected. 

To circumvent the threat of the Trojan attack and the instability of
the interferometers, Walton \textit{et al.} \cite{WASST2003} proposed
a one-way protocol based on decoherence free-subspaces in which each
qubit is encoded in the time and phase of a pair of photons. In this
Letter, we propose a new way to protect qubits encoded in polarization
states of a photon pair from birefringence effects in optical fibre.

The idea is to take advantage of the fact that birefringence can be
well approximated by a collective error model as long as the photons
travel inside a time window small compared to the variation of the birefringence. Thus, if the effect of
birefringence on one photon is $U(t)$, on $n$ photons it is 
$U(t)^{\otimes n}$.  This latter operator can be interpreted as a
rotation of the reference frame axis and our protocol reduces to the
problem of developing a strategy to do quantum communication without a
shared reference frame.

In a recent paper \cite{BRS2003}, Bartlett \textit{et al.} showed it
should be possible to ``{\it communicate with perfect fidelity without a shared reference frame at a rate that asymptotically approaches one encoded qubit per transmitted qubit}.'' In particular, they proposed a method to encode a qubit using four photons in a decoherence-free-subspace of the collective noise model.  However this required having full control of the states of qubits.  This is out of
reach of today's technology. More recently, two realistic QKD
protocols that do not require any shared reference frame have been
proposed \cite{BGLPS2004}. These protocols do not require a general
state of a qubit but only a set of non-orthogonal states.
It encodes qubits in both three and four
photon states, which makes the protocol more sensitive to photon loss.
For these reasons, we
will describe a two photon protocol robust against phase instability
of the interferometer without the need for a shared spatial reference
frame or synchronized clocks. 
If we neglect dispersion and discard
relativistic situations then we are close to having no need for a
shared reference frame at all.\footnote{For reasons we will explain later, Bob needs to know the relative rate of time flow in Alice's reference frame.}

To explain our protocol we need to introduce the ``tag'' operation
$T_{i}$ which delay the photons in the state $\ket i$ by a specific
amount of time.  Experimentally it can be implemented using a
polarized beamsplitter to separate polarization modes in arms of
different length before recombination in the same optical path.

Suppose Alice inputs a two-photon state of the form $\alpha \ket{HV} +
\beta \ket{VH}$ where $H$ and $V$ correspond to the horizontal and
vertical polarization state of a photon. The time delay between the
two photons $\Delta t_p$, must be fixed by Alice and known by Bob. It
must be large enough such that Bob's apparatus can differentiate
between the two photons and that ``tag'' operation will never change
their order of arrival.  If Alice applies the ``tag'' operation $T_V$
on the initial state then she will have
$\alpha \ket{HV_T} + \beta \ket{V_TH}$, where subscript $T$ denotes the delay. 
Suppose some collective noise $U^{\otimes 2}$ (that includes a change of 
reference frame) is applied to this state when it travels to Bob and suppose also that Bob applies the ``tag'' operation $T_{H'}$ when he receives it. Up to a global phase, the state is then mapped to
\begin{eqnarray}
&& \frac{\alpha}{2}(\ket{H'_TV'_T}-\ket{V'H'_{TT}}+\delta_1 (\ket{H'_TV'_T}+\ket{V'H'_{TT}})\nonumber\\
&&+\delta_2 (\ket{H'_TH'_{TT}}+\ket{V'V'_T})+\delta_3 (\ket{H'_TH'_{TT}}-\ket{V'V'_T})  )\nonumber\\
&& +\frac{\beta}{2} (\ket{V'_TH'_T}-\ket{H'_{TT}1'}+\delta_1(\ket{V'_TH'_T}+\ket{H'_{TT}1'})\\ 
&&+\delta_2(\ket{H'_{TT}H'_T}+\ket{V'_TV'})+\delta_3(\ket{H'_{TT}H'_T}-\ket{V'_TV'})) \nonumber
\end{eqnarray} 
where $\ket{H'}$ and $\ket{V'}$ notation is used since the state is
now defined in Bob's reference frame. We used the fact that the anti-symmetric state $\ket{\Psi^{-}}=\frac{1}{\sqrt 2}(\ket{HV} - \ket{VH})$ is invariant under collective noise and that $\ket{\Psi^{+}}=\frac{1}{\sqrt 2}(\ket{HV} + \ket{VH})$ will be mapped to a superposition of the triplet Bell states for which the $\delta$'s represent the relative weights and phases and follow the equality $||\delta_1||^2+||\delta_2||^2+||\delta_3||^2=1$. For later convenience, we define $\ket{\Phi^\pm}=\frac{1}{\sqrt 2}(\ket{HH}\pm\ket{VV})$ and we will drop the apostrophe notation for simplicity.

The last operation is to project onto the states subspace in which the
photons are separated in time by exactly $\Delta t_p$, i.e. both have been subjected to one tag operation. This operation does not require synchronized clocks, since Bob just needs to compare
the arrival time of both photons. If the interval of time between a
pair of photons is not $\Delta t_p$, then he discards these qubits,
which happens $1-||\frac{(1+\delta_1)}{2}||^2$ of the time if we
neglect photon loss. Otherwise, Bob will obtain Alice's initial state
$\alpha \ket{H_TV_T} +\beta \ket{V_TH_T}$ with certainty.
As it could have been showed using simple calculations, the final result is independent of the phase coherence instability between both arms of the interferometer in a way similar to the qubits encoded in the Walton \textit{et al.} protocol \cite{WASST2003}.

To check if the communication is efficient, $||\frac{(1+\delta_1)}{2}||^2$  must be estimated.  If the collective noise is averaged uniformly\footnote{We assume that the randomness of the birefringence is such that the distribution of $U$ over a large amount of time is uniform. The Haar measure over the space of unitary matrices is then used to calculate the average $\langle\bra{\psi} T_{1}^{\dagger} U^{\otimes 2}T_{1}\ket{\psi}\rangle$ which equals $\frac{1}{3}$ independently of $\ket{\psi}$. Consequently, $\langle ||\frac{(1+\delta_1)}{2}||^2\rangle =\frac{1}{3}$.}  over all possible values of $U(t)^{\otimes 2}$, then $\langle
||\frac{(1+\delta_1)}{2}||^2\rangle =\frac{1}{3}$, which means Bob
will obtain Alice's state with a probability of $\frac{1}{3}$. Yet,
this result supposes that the unitary matrix $U$ will average
uniformly over all possible values during the communication time. To
make the protocol independent of the environment, Bob could apply a
random unitary matrix $B^{\otimes 2}$ on the photon polarization
states just before making his ``tag'' operation\footnote{The distribution of the operator $B$ should correspond to the normalized Haar measure.  Experimentally, $B$ could be implemented with Pockels cells the same way as Franson and Jacobs in their 1995 experiment \cite{FJ1995}.}. 
  
An improved version of the scheme exploiting some
partial knowledge of the shared reference frame to modify the
transformation $B$ to approximate the transformation $U^{\dagger}(t)$
would increased the ratio of useful encoded qubits.  Depending on the efficiency of the active
feedback mechanism and the rate of change of $U(t)$, the ratio could
converge to 1.

\begin{figure}[hb!]
 \includegraphics[scale=0.24]{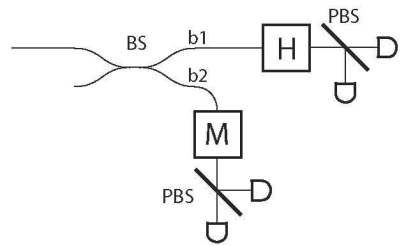}
\caption{\footnotesize{After receiving the two photons and applying his ``tag'' operation, Bob can use this circuit to measure the qubit $\alpha \ket{HV} + \beta \ket{VH}$ in any basis by  adjusting the gate M with a success probability of at least $\frac{1}{8}$. We refer to the text for more details.}}
\label{Circuit1}
\end{figure}

To measure the qubit in a particular basis, Bob could
use a normal symmetric beamsplitter and consider the result when each
photon goes through a different branch, as shown in figure
\ref{Circuit1}. Define $p$ such that $p=0$ if the first photon goes
through branch b1 and 1 if it is the second photon.  Remark that the two photons arrive at the  beamsplitter at different times and that Bob can differentiate them. At the end of
branch b1, Bob measures in his diagonal $\{\ket{+}, \ket{-}\}$
polarization basis. Define $k$ such that $k=0$ if the outcome is
$\ket{+}$ and $1$ if it is $\ket{-}$. The photon on the other branch
b2 must then be in the state $X^{p}Z^{k}(\alpha \ket{H}+\beta
\ket{V})$ where X and Z are the corresponding Pauli operators. Using Pockels cells (M) on the second branch and a polarized beam splitter, Bob can measure the qubit in any specific basis with a chance of success reduced by a factor of at most 8, since at the very least the measurement is successful when each photon exits from a different branch and p=k=0. Measurement in some bases will be successful more often than others.

We have described a technique to encode a robust qubit against 
collective noise and to measure it in any basis. We now show how this could be useful for a realistic QKD implementation. First, we describe the well known QKD protocol BB84 \cite{BB84}. This protocol uses a set of four quantum states consisting of two maximally conjugate basis states $\ket{0}, \ket{1}$ and $\ket{\pm}=\frac{1}{\sqrt{2}}(\ket{0}\pm\ket{1})$. Alice randomly chooses which basis she will use to encode qubits to send Bob, who, upon arrival of a qubit, also chooses at random in which of the two basis he will perform a measurement. After repeating the protocol for a string of random bits, they publicly share what basis they used for each qubit. The bits for which they have used the same basis is used to build the \emph{sifted} key. Since Eve has no prior knowledge of which basis  Alice and Bob will use, any attempt of eavesdropping will disturb the states and induce errors in the \emph{sifted} key with high probability. A portion of the \emph{sifted} key is used to detect possible eavesdropping. If the error rate is lower than some given threshold, the left over bits will be transformed to the final secret key by using error correction and privacy amplification  \cite{M96}.

\begin{figure*}[t!]
\includegraphics[width=4.8 in]{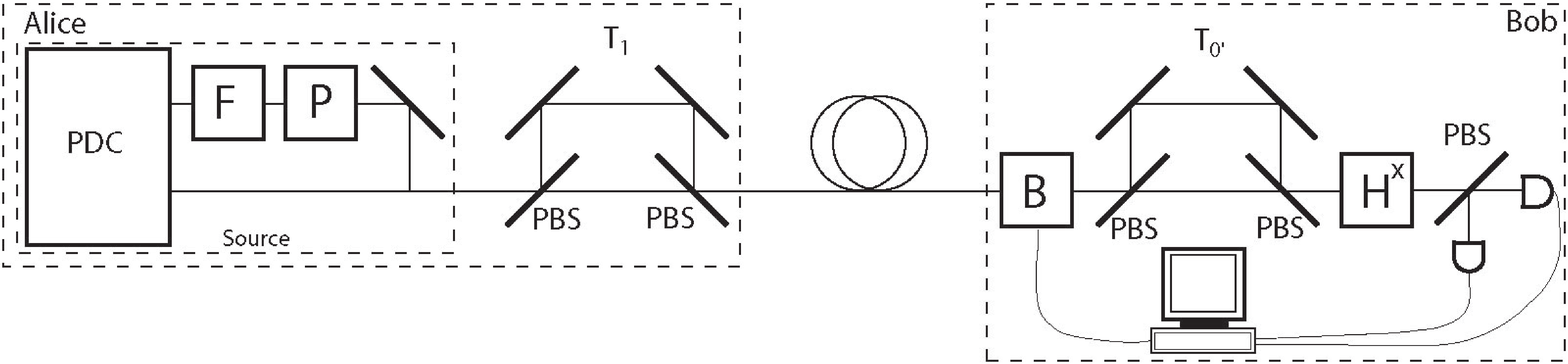}
\caption{\footnotesize{Implementation of a modified version of BB84 protocol based on qubits robust against collective noise. Quantum states are generated through parametric down converters (PDC) supplemented by filters (F) and phase shifter (P). Alice and Bob do their  ``tag'' operation using polarized beamsplitters (PBS). The $B$ operator is randomized uniformly or determined by using a smart feedback mechanism. Bob measure the state in the computational or the diagonal basis depending if he applied the identity ($x=0$) or the Hadamard gate ($x=1$).}}
\label{Circuit2}
\end{figure*}

To implement a protocol similar to BB84, Alice needs to encode the states $\ket{HV_T}$, $\ket {V_TH}$, $\frac{1}{\sqrt 2}(\ket{HV_T} + \ket{V_TH})$ and $\frac{1}{\sqrt 2}(\ket{HV_T} - \ket{V_TH})$ using parametric down conversions, filters and polarized beamsplitters as shown in figure \ref{Circuit2}. We have to note that the measurement procedure described earlier works only if the state received by Bob after post-selection was of the form $\gamma_1 \ket{HV} + \gamma_2 \ket{VH}$ where $\gamma_{i}\in\mathbb{C}$ respecting a normalizing condition. This condition may no longer be true if sources of noise other than collective noise are considered or if we suppose that Eve altered the state sent to Bob. In the latter case, Bob's state after post-selection would look like $\gamma_1 \ket{HV} + \gamma_2 \ket{VH}+\gamma_3 \ket{VV}+\gamma_4 \ket{HH}$. To implement the provenly secure BB84 protocol, Bob must be able to project that state into the subspace in which Alice has encoded her space i.e. the space spanned by $\ket{HV}$ and $\ket{VH}$. If Bob wants to measure in the computational basis ($\{\ket{VH}, \ket{HV}\}$), then immediately after his ``tag''  operation he simply needs to measure the $\ket{H}$ or $\ket{V}$ polarization of each photon. In this case, he will also distinguish and be able to discard the states $\ket{HH}$ and $\ket{VV}$. The measurement in the diagonal basis $\ket{\Psi^{\pm}}$ is not as straight forward. Suppose Bob applies an extra Hadamard gate on both photons before measuring the polarization states. If $\gamma_3 = \gamma_4 =0$, then he measures $\ket{\Psi^{+}}$ if both photons have the same polarization and $\ket{\Psi^{-}}$ if they have different polarization. In general, $\gamma_3 = \gamma_4 \ne 0$, but the uniformly distributed random rotation $B$ performed by Bob (unknown to Eve) when he received the state will destroy any phase coherence between the states  $\gamma_1 \ket{HV} + \gamma_2 \ket{VH}$, $\ket{HH}$ and $\ket{VV}$ from Eve's perspective. Intuitively, this means if Eve used the space spanned by $\{\ket{VV}, \ket{HH}\}$ it would be the same as if she randomly sent one of $\ket{\Psi^{-}}$ or $\ket{\Psi^{+}}$ to Bob, giving her no advantage. The complexity of the QKD security proof which includes coherent 
attacks restrains our argument, but the authors conjuncture that our protocol is unconditionally secure with the same error threshold as BB84. As a last remark, we note that only the qubits that have survived the post-selection are used to build the \emph{sifted} key to estimate the error rate and construct the final secret key.

Earlier we discussed the possibility of using a feedback mechanism to increase the success rate of the post-selection. It could also be used in the QKD implementation discussion above, but Bob must be careful with whatever mechanism he uses since he must ensure the phase coherence between the three states $\gamma_1 \ket{HV} + \gamma_2 \ket{VH}$, $\ket{HH}$ and  $\ket{VV}$ be lost from Eve's perspective. A final random phase gate would be enough since it does not affect the success probability of the post selection, but will destroy the coherence between these states.

The advantages of our
protocol over the plug -and-play one are that this protocol is one-way, so
there is no need to be as worried with the Trojan attack. Moreover, it does not require
interferometer stability like in the Walton \textit{et al.} protocol (by using decoherence-free subspace). 
Although our protocol has similarities to the latter protocol, it is distinct for the following reasons:

First, synchronized clocks are necessary in our protocol only to label the different photon pairs. In the Walton \textit{et al.} protocol, Bob
must be able to distinguish between photons that have been delayed
once, twice and not at all. Our protocol just needs to compare the delay
between the two photons and not their particular time of arrival. Consequently, it requires a much smaller order of timing precision. For example, parametric down-conversion sources with long
pulse length no longer induce errors caused by uncertainty in the
emission time since both photons are always created
simultaneously. Remark that if the number of events in which simultaneous dark counts on different detectors occur is negligible, extra timing precision would not help Alice and Bob to reduce the noise caused by the detector's dark counts and is therefore not necessary to our protocol.

Second, in the Walton \textit{et al.} protocol, there is a
$\frac{1}{4}$ chance, independent of the birefringence, that the
photons will be measured in the phase basis and a $\frac{3}{4}$ chance
of measuring in the time basis. However, the optimal efficiency for
the ideal implementation of BB84 is a probability of measurement equal to $\frac{1}{2}$ in each basis. For this reason, Walton \textit{et al.}
indicate that the intrinsic efficiency of their scheme was
$\frac{1}{4}$. In the case where $B$ is chosen from a uniform distribution, our protocol would
have an intrinsic efficiency ratio of $\frac{1}{6}$ since only a third of the photon pairs is not discarded.  However, depending on the feedback mechanism, the intrinsic efficiency ratio could be higher than $\frac{1}{6}$, up to
$\frac{1}{2}$.

Third, the final state Bob uses is encoded in polarization, not in time and phase. A good control of the polarization states allows Bob to get ride of the noise caused by the polarization dependance of some experimental components, like phase modulators.

In this paper, we have given a realistic robust scheme to do single qubit communication using two-photon states per encoded qubit. This technique goes around the problem of birefringence in optical fibre, the requirement of high precision synchronized timing and also the interferometer phase coherence instability. The protocol could be slightly modified to exploit partial information about a spatial reference frame to increase the bit rate by using active feedback. We also explained how to implement a slightly modified version of BB84 using the previously mentioned methods.

We would like to conclude with some problems that could make an experimental implementation of our schemes more difficult. Depolarization could be a serious distance limitation for our protocol, forcing us to use sources with longer coherence times \cite{GRTZ2002}. To prevent chromatic dispersion from affecting the time delays between the photons, the average wavelength of the photons should be chosen according to the zero chromatic dispersion of the optical fiber \cite{GRTZ2002, F1992}. Finally, since our protocol encoded each qubit with two photons, attenuation and detector's inefficiencies have a more significant affect on its efficiency compared to one-photon protocols. Nevertheless our proposal is in reach of experimental implementation and provides an elegant solution to the problem of birefringence in optical fibres.

We wish to thank Konrad Banaszek, Joseph Emerson, Simon Leblond and Rob Spekkens for
helpful discussions that inspired this work. J.C.B., M.L., R.L.
acknowledge support NSERC and R.L. from ARDA.


\begin{thebibliography}{99} 

\bibitem{qc1983} Wiesner, S., {\it Sigact news}, 15:1, 78-88 (1983).

\bibitem{BB84} C. H. Bennett and G. Brassard, in {\it Proceeding of the IEEE International Conference on Computers, Systems, and Signal Processing, Bangalore}, India (IEEE, New York, 1984), pp.175-179 (1984).

\bibitem{BBBSJ1992} C. H. Bennett, F. Bessette, G. Brassard, L. Salvail, and J. Smolin, {\it J. of Cryptology} {\bf 5} 3 (1992). 

\bibitem{GRTZ2002} N. Gisin, G. Ribordy, W. Tittel, and H. Zbinden, {\it Rev. Mod. Phys.} {\bf 74} 145 (2002).

\bibitem{GBGGHRTZ2000} N. Gisin, J. Brendel, J-D. Gautier, B. Gisin, B. Huttner, G. Ribordy, W. Tittel and H. Zbinden {\it Lect. Notes Phys.} {\bf 538} 191-200 (2000).

\bibitem{FJ1995} J.D. Franson and B.C. Jacobs, {\it Electron. Lett.}Ê {\bf 31} 232-234 (1995). 

\bibitem{J1991}  K.R.W. Jones, {\it Annals of Phys.} {\bf 207} 140-147 (1991).

\bibitem{RG2003} T. Rudolph and L. Grover, {\it Phys. Rev. Lett.} {\bf 91} 217905 (2003).

\bibitem{Bennett1992} C.H. Bennett, {\it Phys. Rev. Lett.} {\bf 68}, 3121 (1992).

\bibitem{TRT1993} P. Townsend, J.G. Rarity and P.R. Tapster, {\it Electron. Lett.} {\bf 29} 634 (1993).

\bibitem{HMP2000} R.J. Hughes, G.L. Morgan and C.G. Peterson, {\it J. of Mod. Opt.} {\bf 47} 533-547 (2000).

\bibitem{MHHTZG1997}  A. Muller, T.Herzog, B. Huttner, W.Tittel, H. Zbinden and N. Gisin, {\it Applied Phys. Lett.} 70 (7) 793-795 (1997)

\bibitem{SGGRZ2002} D. Stucki, N Gisin, O. Guinnard, G. Ribordy and H. Zbinden, {\it New J. Phys.} {\bf 4} 41 (2002). 

\bibitem{Martinelli1989} M. Martinelli, {\it Opt. Commun.} {\bf 72} 341 (1989).

\bibitem{WASST2003} Z.D. Walton, A.F. Abouraddy, A.V. Sergienko, B.E.A. Saleh and M.C. Teich,  {\it Phys. Rev. Lett.} {\bf 91} 087901 (2003).

\bibitem{BRS2003} S. D. Bartlett, T. Rudolph and R. W. Spekkens, {\it Phys. Rev. Lett.} {\bf 91} 027901 (2003).

\bibitem{BGLPS2004} J.-C. Boileau, D. Gottesman, R. Laflamme, D. Poulin and R. W. Spekkens, {\it Phys. Rev. Lett.} {\bf 92} 017901 (2004).

\bibitem{M96} D. Mayers, in {\it Advances in Cryptology: Proceedings of CryptoÕ96}, Lecture Notes in Computer Science Vol. 1109 (Springer-Verlag, Berlin, 1996), p. 343;  P.W. Shor and J. Preskill, {\it Phys. Rev. Lett.} {\bf 85} 441-444 (2000).


\bibitem{F1992}  J.D.Franson, {\it Phys. Rev. A} {\bf 45} 3126-3132 (1992);  A.M. Steinberg, P. Kwiat and R.Y. Chiao,  {\it Phys. Rev. A} {\bf 45} 6659-6665 (1992).


\end{thebibliography}
\end{document}